\documentclass[aps,pra,superscriptaddress,twocolumn]{revtex4-2}
\usepackage[utf8]{inputenc}
\usepackage{amsmath, bm}
\usepackage{amsfonts}
\usepackage{mathtools}
\usepackage{physics}
\usepackage{dsfont}
\usepackage{verbatim}
\usepackage{color}
\usepackage[dvipsnames]{xcolor}

\usepackage[normalem]{ulem}

\usepackage[normalem]{ulem}
\usepackage{soul}
\usepackage[utf8]{inputenc}
\usepackage[ruled,longend]{algorithm2e}
\usepackage{textgreek}
\usepackage{amssymb}
\usepackage{eucal}
\usepackage{tcolorbox}
\usepackage{wasysym}

\def\Var{\mathop{\rm Var}} 

\newcommand\smallO{
  \mathchoice
    {{\scriptstyle\mathcal{O}}}
    {{\scriptstyle\mathcal{O}}}
    {{\scriptscriptstyle\mathcal{O}}}
    {\scalebox{.7}{$\scriptscriptstyle\mathcal{O}$}}
  }

\begin{document}
\title{Engineered dissipation to mitigate barren plateaus}

\author{Antonio Sannia}
\affiliation{%
 Institute for Cross-Disciplinary Physics and Complex Systems (IFISC) UIB-CSIC, Campus Universitat Illes Balears, 07122, Palma de Mallorca, Spain.
}%

\author{Francesco Tacchino}
\affiliation{IBM Quantum, IBM Research Europe – Zurich, S\"aumerstrasse 4, 8803 R\"uschlikon, Switzerland}

\author{Ivano Tavernelli}
\affiliation{IBM Quantum, IBM Research Europe – Zurich, S\"aumerstrasse 4, 8803 R\"uschlikon, Switzerland}
\author{Gian Luca Giorgi}
\affiliation{%
 Institute for Cross-Disciplinary Physics and Complex Systems (IFISC) UIB-CSIC, Campus Universitat Illes Balears, 07122, Palma de Mallorca, Spain.
}%

\author{Roberta Zambrini}%
\affiliation{%
 Institute for Cross-Disciplinary Physics and Complex Systems (IFISC) UIB-CSIC, Campus Universitat Illes Balears, 07122, Palma de Mallorca, Spain.
}%

\begin{abstract}
Variational quantum algorithms represent a powerful approach for solving optimization problems on noisy quantum computers, with a broad spectrum of potential applications ranging from chemistry to machine learning. However, their performances in practical implementations crucially depend on the effectiveness of quantum circuit training, which can be severely limited by phenomena such as barren plateaus. While, in general, dissipation is detrimental for quantum algorithms, and noise itself can actually induce barren plateaus, here we describe how the inclusion of properly engineered Markovian losses after each unitary quantum circuit layer can restore the trainability of quantum models. We identify the required form of the dissipation processes and establish that their optimization is efficient. We benchmark our proposal in both a synthetic and a practical quantum chemistry example, demonstrating its effectiveness and potential impact across different domains.
\end{abstract}

\maketitle

\section{Introduction}

While dissipation is generally detrimental to quantum technologies and is, in fact, a limiting factor for currently available noisy quantum devices~\cite{preskill2018quantum}, there are important exceptions. On the one hand, going beyond unitary operations is not only needed to account for quantum measurements \cite{Wiseman2009} but also enables several applications ranging from quantum state preparation and control to error correction \cite{DissipEng1}. On the other hand, noisy quantum platforms are well suited for computing \cite{bharti2022noisy} in a hybrid quantum-classical setting, for instance using the popular class of variational quantum algorithms (VQAs) \cite{cerezo2021variational}. In this framework, a parametric quantum circuit is used to prepare complex quantum states and to evaluate, through sampling or with suitable quantum measurements, a cost function that would be otherwise expensive to compute classically. The optimization of the circuit parameters, aimed at minimizing such cost, is instead assigned to a classical optimizer. As an example, variational quantum eigensolvers (VQEs) \cite{Peruzzo2014,tilly2022variational} can be used to approximate the ground state of a given Hamiltonian \textit{H} through a trial quantum circuit, also called an ansatz, by minimizing the cost function given by the expectation value of \textit{H}. More in general, VQAs can be applied to classical optimization problems~\cite{Moll2018}, linear systems of equations~\cite{vqa_lin, vqa_lin2, vqa_lin3}, quantum simulations~\cite{vqa_qsim, vqa_qsim2, vqa_qsim3, vqa_qsim4}, quantum data compression~\cite{qd_compression}, 
quantum machine learning~\cite{havlivcek2019supervised,Benedetti_2019,Mangini_2021}, 
generative models~\cite{VQA_gen1, VQA_gen2, VQA_gen3}, 
quantum foundations~\cite{q_foundations}, 
quantum compiling~\cite{VQA_compiling, VQA_compiling2, VQA_compiling3, VQA_compiling4}, and quantum error correction~\cite{VAQ_errorcorrection}. 
In many of these cases, the VQA can also be formulated as a ground-state problem with a problem-specific (rather than a physically motivated) Hamiltonian, the choice of which is generally non-trivial~\cite{cerezo2021variational}.

Despite their potential advantages, VQAs are known to suffer from a number of bottlenecks that still prevent their implementation at scale. One particularly serious drawback, specific to quantum operations, is the phenomenon known as barren plateaus~\cite{mcclean2018barren}. These manifest as a vanishing gradient of the loss function with respect to the model parameters and can lead to severe limitations in the training efficiency. Barren plateaus build up as the system size increases, hence directly hindering the scalability of VQAs. Specifically, it was found that when a quantum circuit ansatz reaches the 2-design limit the probability of randomly inducing a non-negligible gradient decreases exponentially with the number of qubits~\cite{mcclean2018barren}. This condition is relatively easy to satisfy~\cite{2design, 2design2, 2design3} implying a generalized loss of all possible quantum speedups, even if the optimization strategy does not include gradient computation \cite{gradient_free}. 

Recent findings indicate that barren plateaus can originate from several factors, including high circuit expressibility~\cite{expressibility, larocca2022diagnosing}, concentration of the cost function~\cite{cost_conc}, noise~\cite{NoiseBP},  entanglement excess~\cite{entang}, and globality of the cost function~\cite{cost_func, local_Uvarov2021}. All these barren-plateau sources can be unified by means of a Lie algebra theory that applies to a general (yet unitary) setting~\cite{formula_BP1,formula_BP2}.
Several techniques to mitigate barren plateaus have been reported, such as initialization strategies~\cite{mit_initialization, CDNN}, transferability of smooth solutions~\cite{Mele}, entanglement limitations~\cite{mit_entang}, correlations and restrictions of circuit parameters~\cite{expressibility, corr_par}, classical shadows~\cite{classical_shadows}, pre-trainings~\cite{pre-training, pre_train2}, and layerwise learning~\cite{layer,Tacchino2020IEEE}. These mitigation strategies generally consist in appropriately constraining the unitary ansatz. In this work, we change the focus acting on the problem Hamiltonian without adapting the quantum circuit. Remarkably, the occurrence of barren plateaus is closely related to the unitarity of the ansatz \cite{formula_BP1, formula_BP2}, while assessing these phenomena beyond a fully unitary framework presents a promising research avenue. Our goal is therefore to demonstrate the potential of non-unitary, open-system dynamics as a powerful strategy to overcome the trainability barrier and to ensure efficient convergence. To achieve it, it is reasonable to expect that a non-trivial engineering of the dissipation processes will be required. Indeed, while it has been already suggested that non-unitary operations can increase the accuracy of ground-state VQE calculations in quantum chemistry ~\cite{nuVQE_PRL,NU-VQE}, generic noise is known to actually induce barren plateaus~\cite{NoiseBP,Melo2023pulseefficient}. In parallel, dissipation implemented simply by discarding qubit registers, as in the class of so-called dissipative quantum neural network models \cite{training_DQNN, quantum_autoencoders, no_freelunch}, is also not sufficient to ensure trainability, in general~\cite{training_DQNN}. The proof that a suitable set of operations can in fact be constructed constitutes the main technical contribution of our work.

Building on previous results obtained in the field of engineered quantum dissipation \cite{verstraete2009quantum, pastawski2011quantum}, we consider a Markov dissipation modeled by a Gorini-Kossakowski-Sudarshan-Lindblad (GKLS) Master Equation \cite{breuer2002theory, gorini1976completely, lindblad1976generators}. Our strategy to design the proper dissipation is motivated by the recent observation that training a \textit{local} cost function (a cost function made of local observables) is much more efficient than training a \textit{global} one \cite{emp_loc1, emp_loc2, emp_loc3,letcher2023tight}. Rigorous analyses have proven that local cost functions, unlike global ones, are immune to barren plateaus in the case of shallow circuits \cite{local_Uvarov2021, cost_func}. While formulating variational algorithms locally is preferable, mapping a global problem to a local one is generally a non-trivial task. Here, we propose variational algorithms based on non-unitary ansatzes and we show that this represents an effective strategy to tackle barren plateaus (Sec. \ref{Sec:Non_Un}), setting the theory framework (Sec. \ref{Sec:main}) and discussing both a synthetic and a chemical example (Sec. \ref{Sec:Rand_Ham} and Sec. \ref{Sec:Chem_Ex}).

 \section{General framework}\label{Sec:VQE_Unitary}

\begin{figure*}[]
    \includegraphics[width= \linewidth, keepaspectratio]{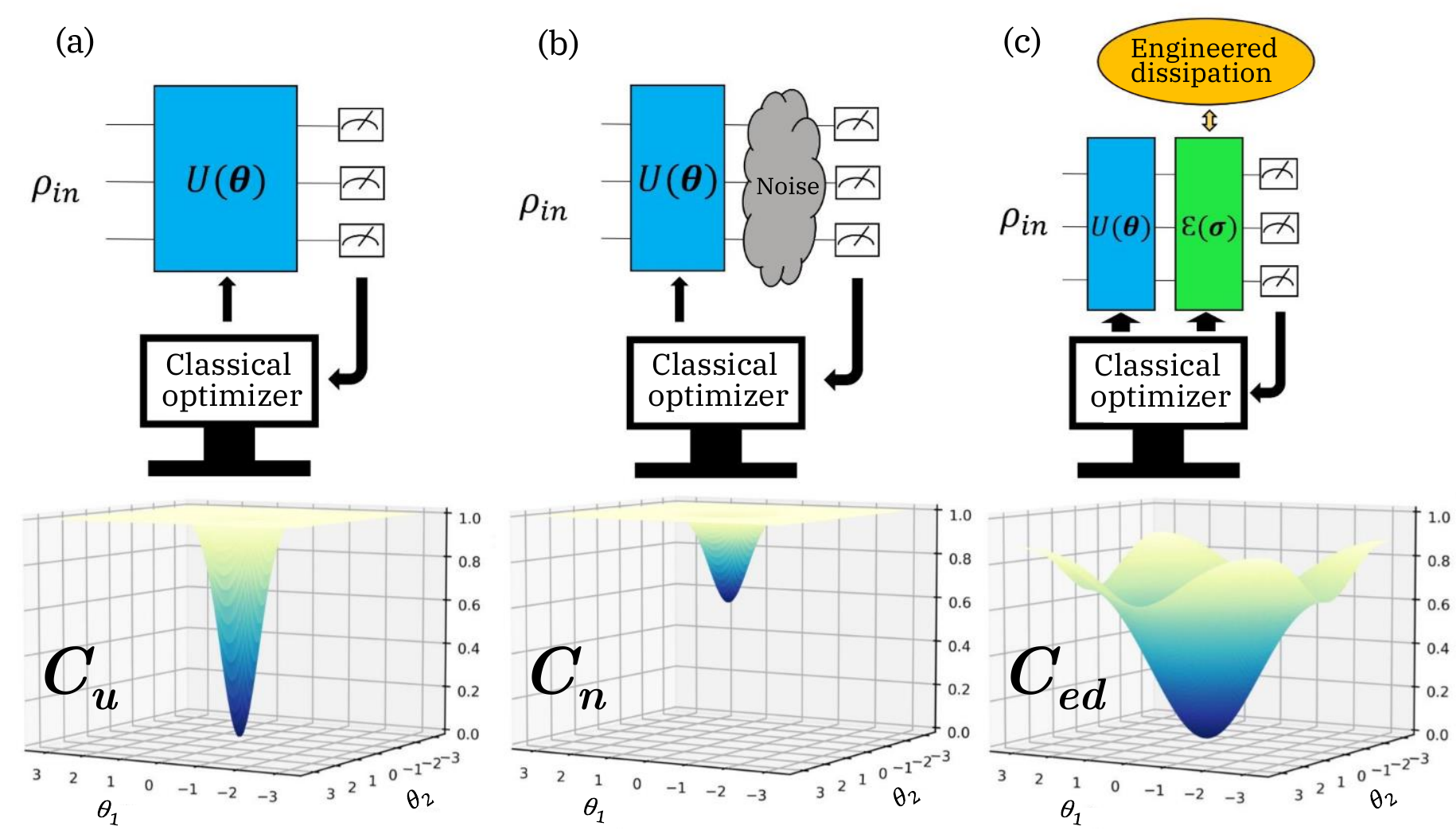}
    \caption{Cross sections of the warm-up example cost function landscapes of Sec. \ref{Sec:WarmUp} for a 20-qubit system. (a) Fully unitary ansatz $C_u$, (b) noisy landscape $C_n$ with depolarizing probability $p = 0.5$, and (c) engineered-dissipation ansatz $C_{ed}$ in the case of   $\Delta t \simeq 2,33$, which corresponds to the maximum of $\Var[\partial C_{ed}/\partial \theta_j]$ (see main text).}
    \label{fig:Landscapes}
\end{figure*}

In the most general case, VQAs consist of minimizing a cost function whose minimum faithfully corresponds to the solution of a considered problem \cite{cerezo2021variational}. In the following, we consider the case where the cost function corresponds to the expectation value of an $n$-qubit Hermitian operator $\textit{H}$ and, consequently, the solution is its ground-state energy. Given an initial condition $\rho_{in}$ and a quantum circuit ansatz $U(\bm{\theta})$, where $\bm{\theta}$ is a free parameter vector, the cost function has the form
\begin{align*}
    C(\bm{\theta})=\Tr\{H  U(\bm{\theta}) \rho_{in} U^{\dagger}(\bm{\theta})\}.
\end{align*}
The minimization strategy consists of evaluating $C(\bm{\theta})$ by a quantum device. The free parameters are optimized by a classical procedure.

A necessary condition for the circuit trainability is that, for a random initialization of $\bm{\theta}$, the probability of finding a non-negligible value of the cost function derivative with respect to a generic parameter $\theta_k$, denoted as $\partial_{k}C$, is itself not negligible.
An upper bound on this probability is known to be proportional to the variance of the derivative, which we write as $\Var[\partial_{k}C]$, setting the limits to training efficiency \cite{mcclean2018barren,expressibility, larocca2022diagnosing, cost_conc, NoiseBP, entang, cost_func, local_Uvarov2021, formula_BP1,formula_BP2}. By definition, we say that the cost function landscape presents a barren plateau if $\Var[\partial_{k}C]$ exponentially decreases with the number of qubits $n$, i.e., $\Var[\partial_{k}C] \in \mathcal{O}(e^{-pn})$ where $p$ is a positive integer.

This phenomenon strongly depends on the locality of $H$ where locality, in this context, refers to the number of qubits on which the Hamiltonian acts non-trivially. We expand the Hamiltonian such that
\begin{align}\label{Eq:Local_H}
H = c_0 \mathbb{I} + \sum_{i=0}^{N} c_i H_i
\end{align}
where $\mathbb{I}$ is the $n$-qubit identity operator, $H_i$ is a generic Hermitian operator and $c_i$ is a real coefficient. By definition, we say that $H$ is \textit{local} when all $H_i$ terms act non-trivially on at most $K$ qubits (where $K$ does not scale with $n$), as is the case of nearest-neighbor interaction Hamiltonians. On the other hand, we call $H$ \textit{global} if the $H_i$ operators act non-trivially on all qubits. It has been rigorously proved that, under quite general conditions, for \textit{alternating layered ansatzes} of arbitrary depth, a barren plateau is inevitable in the global case, while in the local case shallow circuits can prevent it~\cite{cost_func,local_Uvarov2021}. More precisely, if the number of layers (L) does not increase faster than a logarithm in the number of qubits, i.e., $L = \mathcal{O} (\log(n))$ then:
\begin{align*}
    \Var[\partial_{k}C] = \Omega \left(\frac{1}{\text{poly}(n)} \right)
\end{align*}
which means that the probability of finding a non-negligible gradient does not decrease faster than a polynomial implying the absence of barren plateaus.

It has also been empirically found that when a problem can be encoded with both a global and a local cost function, the training is more effective in the local case and the mitigation techniques work much better \cite{expressibility,emp_loc1, emp_loc2, emp_loc3}. Moreover, it has been also analytically shown that $\Var[\partial_{k}C]$, in general,  increases with the Hamiltonian locality \cite{formula_BP1,formula_BP2}. In the following, we will focus on the case where the Hamiltonian of the problem is originally global and an equivalent local one is unknown. We will show how the addition of a proper non-unitary layer in the variational scheme allows the problem to be approximated with a local one where barren plateaus are absent or easier to face.

\section{Non-unitary Ansatz}\label{Sec:Non_Un}

We propose to generalize the unitary framework for VQA to Markovian maps, designing the proper dissipation \cite{verstraete2009quantum, pastawski2011quantum} amenable to experimental implementation, also in near-term quantum devices. Previous non-unitary ansatz in VQA's 
have been considered for \textit{in silico} implementations (classical post-processing) to boost precision in chemical problems \cite{nuVQE_PRL,NU-VQE}. The non-unitary ansatz acting on the quantum state is
\begin{align*}
    \Phi(\bm{\sigma}, \bm{\theta}) \rho = \mathcal{E}(\bm{\sigma}) \circ U(\bm{\theta}) \rho U^{\dagger}(\bm{\theta})
\end{align*}
where $U(\bm{\theta})$ is, as before, a parametric quantum circuit, and $\mathcal{E}(\bm{\sigma})$ is a non-unitary superoperator, with their respective tunable parameters $\bm{\theta}$ and $\bm{\sigma}$. Under Markovian assumption, the non-unitary part of the ansatz is defined by a parametric Liouvillian $\mathcal{L} = \mathcal{L}(\bm{\sigma})$ such that
\begin{align*}
    \mathcal{E}(\bm{\sigma}) = e^{\mathcal{L}(\bm{\sigma})\Delta t},
\end{align*}
where $\Delta t$ is the interaction time with the environment. 
The Liouvillian $\mathcal{L}$ is the superoperator that generates the dynamics of the GKLS Master Equation \cite{breuer2002theory, gorini1976completely, lindblad1976generators}:\begin{equation} \label{eq:Linb}
\dot{\rho}=\mathcal{L}\rho \equiv -i[\mathfrak{H},\rho]+\sum_{i}\gamma_{i} ( L_{i}\rho L_{i}^{\dagger}-\frac{1}{2} \{L_{i}^{\dagger}L_{i},\rho \})
\end{equation}
where $\mathfrak{H}$ is the Hamiltonian responsible of the unitary part of the evolution, $\{\gamma_{i}\}$ are the damping rates and the operators $\{L_{i}\}$, called jump operators, identify the environment action on the qubits.

For our specific proposal, we set the following conditions on $\mathcal{L}$:
\begin{enumerate}
\item $\mathcal{L}$ can be expressed as the sum of $Q$ superoperators $\mathcal{L}_q(\bm{\sigma}_q)$, each of which acts non-trivially on at most $K$ qubits. This expansion takes the form:
\begin{align}\label{Eq:local_dec}
\mathcal{L}(\bm{\sigma}) = \sum_{q=1}^{Q} \mathcal{L}_q(\bm{\sigma}_q).
\end{align}
\item All the generators $\mathcal{L}_q$ commute with each other.
\item All the generators $\mathcal{L}_q$ have exactly one stationary state, $\rho_{ss, q}$.
\item The generators $\mathcal{L}_q$ converge to their respective stationary states at the same rate, which we refer to as the mixing time.
\end{enumerate}

According to the previous definitions {and under the assumptions 1. and 2., the cost function reads}: 
\begin{align}
    C(\bm{\theta}, \bm{\sigma}) &=\Tr{H e^{\mathcal{L}(\bm{\sigma}) \Delta t} U(\bm{\theta}) \rho_{in} U^{\dagger}(\bm{\theta})} \nonumber \\
    &=\Tr{H \prod_{q=1}^{Q} e^{\mathcal{L}_q(\bm{\sigma}_q) \Delta t} U(\bm{\theta}) \rho_{in} U^{\dagger}(\bm{\theta})}.\label{Eq:Cost_NU}
\end{align}

\subsection{Illustrative example}\label{Sec:WarmUp}

Before delving into the formal characterization, we present a simple example to illustrate the strategy of tackling the barren plateaus with non-unitary VQAs. Following \cite{cost_func}, we consider a state preparation task formulated through the optimization of a global Hamiltonian. In particular, we are interested in minimizing $H = \mathbb{I} - \ket{\bm{0}}\bra{\bm{0}}$ supposing that all the qubits are initialized in state $\ket{0}$. By applying the unitary ansatz $U(\bm{\theta}) = \bigotimes_{j=1}^{n} e^{- i \theta_j \sigma_x / 2}$, we can easily detect the occurrence of a barren plateau (see Fig. \ref{fig:Landscapes} a). In fact, the cost function takes the form $C_u = 1 - \prod_{j=1}^{n} \cos^{2}( \frac{\theta_j}{2})$ and a direct calculation shows that $ \Var[\frac{\partial C_u}{\partial \theta_j}] = \frac{1}{8} (\frac{3}{8})^{n-1}$. In addition, the derivative values are unbiased, i.e. $ \left\langle\frac{\partial C_u}{\partial \theta_j} \right\rangle = 0$, implying an exponential suppression of the probability to sample a non-negligible gradient as a consequence of Chebyshev’s
inequality.

We now emphasize that, as shown in \cite{NoiseBP}, general noisy non-unitary interactions are not able to mitigate barren plateaus and can actually induce them \cite{NoiseBP}. For example, if we model the noise effect with a depolarizing channel of probability $p$, the resulting cost function will be $C_{n} = p ( 1 - \frac{1}{2^{n}}) + (1-p) C_{u}$  (see Fig. \ref{fig:Landscapes} b). This non-unitary model, in addition to clearly having the same trainability problems of the cost function $C_{u}$, also precludes the possibility of finding the correct ground state energy. 

We now introduce an engineered dissipation through a non-unitary layer, ensuring that each $\mathcal{L}_q$ operator acts dissipatively on a single qubit as per Eq. (\ref{eq:Linb}). We assume that, in general, the $\mathcal{L}_q$ structure lacks a Hamiltonian part and instead is determined by a single jump operator $L_q = \ket{0}_q\bra{1}$  with a corresponding damping rate normalized to $1$. This situation is advantageous because the stationary state of the whole Liouvillian is the ground state of the sought-after problem. The cost function in the presence of such an engineered dissipation is now $C_{ed} = 1 - \prod_{j=1}^{n}[1 - \sin^{2} (\frac{\theta_j}{2}) e^{-\Delta t}]$ for which the unbiased condition on the gradient still holds. However, in this case, it is possible to efficiently avoid the barren plateau (see broader and still deep minimum in Fig. \ref{fig:Landscapes} c). In fact, $\Var[\frac{\partial C_{ed}}{\partial \theta_j}] = \frac{1}{8} e^{- \Delta t} (1+\frac{3}{8}e^{-2 \Delta t} - e^{-\Delta t})^{n-1}$ and if $\Delta t \sim \mathcal{O} (\log(n))$ the desired polynomial scaling is achieved.

We observe that the $C_{ed}$ landscape takes a similar shape as the one in \cite{cost_func}, where the unitary ansatz was the same as we have considered here, but the global Hamiltonian was replaced by an equivalent local one. 
In the following, we will show how this landscape similarity is not a coincidence because the considered non-unitary ansatz allows, in general, localizing a problem that has been originally formulated as a global one. Moreover, we will also prove that the logarithmic $\Delta t$ scaling is a general feature of the procedure.

\section{Theoretical  results}\label{Sec:main}
Now we will show how a non-unitary layer that respects the constraints introduced in Sec. \ref{Sec:Non_Un} is able to make the Hamiltonian of the cost function local.  First, we note that the diagonalizability of the $\mathcal{L}_q$ superoperators is a mild condition that can be assumed to hold. In fact, if $\mathcal{L}_q$ has a holomorphic dependence on the parameters $\bm{\sigma}_q$ and if it is diagonalizable for a subspace of them,  then it will be diagonalizable in general, apart from the possible appearance of countable exceptional points, which can be ignored in the discussion \cite{Spectral_liov}. Anyway, in all the cases discussed in the following,  diagonalizability will always be ensured. Consequently, it is useful to expand the $\mathcal{L}_q$ superoperators through their dual basis. To keep the notation simple, we will assume that each $\mathcal{L}_q$ acts non-trivially on exactly $K$ qubits \footnote{The following calculations can be immediately generalized to cases where each $\mathcal{L}_k$ operator acts on a generic number of qubits less than or equal to $K$.}. Introducing the Liouville notation for a generic matrix, $\rho \rightarrow | \rho \rangle \! \rangle$, and considering the dot product  $\langle \! \langle \tau | \rho \rangle \! \rangle = \Tr{\tau^{\dagger} \rho}$, then
\begin{align}\label{Eq:Expansion}
    e^{\mathcal{L}_q(\bm{\sigma}_q) \Delta t} = \sum_{i=0}^{4^{K}-1} e^{\lambda_{q, i} \Delta t} \frac{| r_{q, i} \rangle \! \rangle   \langle \! \langle l_{q, i} |}{ \langle \! \langle l_{q, i} | r_{q, i} \rangle \! \rangle },
\end{align}
where $\{\lambda_{q, i}\}_i$ is the set of the eigenvalues of $\mathcal{L}_{q}$ and $\{| r_{q, i} \rangle \! \rangle\}_i$ and $\{| l_{q, i} \rangle \! \rangle\}_i$ are the corresponding set of orthogonal and normalized right and left eigenvectors \footnote{we have omitted the $\bm{\sigma}_q$ dependence of all these terms to make the notation easier}. Ordering the indexes as a function of the real parts of the eigenvalues such that $\Re{\lambda_{q, 0}} > \Re{\lambda_{q, 1}} \ge \dots \ge \Re{\lambda_{q, 4^K-1}}$, and assuming the uniqueness of the steady state $|\rho_{ss, q} \rangle \! \rangle$ (this is also a mild condition to satisfy), we identify $\lambda_{0}=0$, $| r_{q, 0} \rangle \! \rangle = |\rho_{ss, q} \rangle \! \rangle$ and $| l_{q, 0} \rangle \! \rangle = | \mathbb{I} \rangle \! \rangle $.

Equation \ref{Eq:Expansion} can be used to calculate the order of magnitude of the mixing times $\Delta t_{mix, q}$. Indeed, excluding the occurrence of singular phenomena such as the skin effect \cite{PhysRevLett.127.070402}, we have
$\Delta t_{mix, q} \sim \mathcal{O}(1/|\Re{\lambda_{q, 1}}|)$, where the quantity $|\Re{\lambda_{q, 1}}|$ represents the spectral gap. Our constraint of a common mixing time for all $\mathcal{L}_q$ is, then, satisfied if the spectral gaps are equal and do not vary with $\bm{\sigma}_q$. Then, we deal with a unique mixing time, which we will refer to as $\Delta t_{mix}$.

Defining $| \rho(\bm{\theta}) \rangle \! \rangle \equiv | U(\bm{\theta}) \rho_{in} U^{\dagger}(\bm{\theta}) \rangle \! \rangle$, we can rewrite Eq. (\ref{Eq:Cost_NU}) with the introduced notation:
\begin{align*}
    &C(\bm{\theta}, \bm{\sigma}) = \langle \! \langle H | e^{\mathcal{L}(\bm{\sigma}) \Delta t} | \rho(\bm{\theta}) \rangle \! \rangle \\
    &= \langle \! \langle H | \sum_{i_1, \dots, i_Q = 0}^{4^K -1} e^{(\sum_{q=1}^{Q} \lambda_{q,i_q}) \Delta t} \bigotimes_{q=1}^{Q} \frac{| r_{q, i_q} \rangle \! \rangle   \langle \! \langle l_{q, i_q} |}{ \langle \! \langle l_{q, i_q} | r_{q, i_q} \rangle \! \rangle } | \rho(\bm{\theta}) \rangle \! \rangle.
\end{align*}
For a value of the time such that $\Delta t \apprle \log(Q) \cdot \Delta t_{mix} $ an approximation of the non-unitary superoperator, in which only the slowest decay terms are taken, is justified (Theorem 6 in \cite{mixing_time}): 
\begin{align}\label{Eq:Approx_Mark}
    &e^{\mathcal{L}(\bm{\sigma}) \Delta t} = \bigotimes_{q=1}^{Q} \frac{| \rho_{ss, q} \rangle \! \rangle   \langle \! \langle \mathbb{I} |}{ \langle \! \langle \mathbb{I} | \rho_{ss, q} \rangle \! \rangle } \nonumber \\
    &+ \sum_{q=1}^{Q} e^{\lambda_{q, 1} \Delta t} \frac{| r_{q, 1} \rangle \! \rangle   \langle \! \langle l_{q, 1} |}{ \langle \! \langle l_{q, 1} | r_{q, 1} \rangle \! \rangle } \bigotimes_{j \ne q}^{Q} \frac{| \rho_{ss, j} \rangle \! \rangle   \langle \! \langle \mathbb{I} |}{ \langle \! \langle \mathbb{I} | \rho_{ss, j} \rangle \! \rangle } \nonumber \\
    & + \smallO \left(e^{-\Delta t / ( \log(Q) \cdot \Delta t_{mix})}\right).
\end{align}
Moving to the Heisenberg picture, the variational problem, defined by the cost function of Eq. (\ref{Eq:Cost_NU}), can be formulated as a unitary optimization on an effective Hamiltonian $H^{'}$ such that: 
\begin{align} \label{Eq:Heis_pict}
     H^{'}  =  e^{\mathcal{L^{\dagger}}(\bm{\sigma}) \Delta t} H.
\end{align}
Writing $H$ as a linear combination of the $\mathcal{L}$ left eigenvectors (which form a complete set)
\begin{align*}
    H  =\sum_{j_1,\dots, j_Q} c_{j_1, \dots, j_Q} \bigotimes_{q=1}^{Q} | l_{q, i_q} \rangle \! \rangle 
\end{align*}
and assuming  that higher order terms in Eq. (\ref{Eq:Approx_Mark}) can be disregarded, we arrive at the following approximate expression for $H^{'}$: 
\begin{align}\label{Eq:H'}
    H^{'} \simeq  c'_{0}(\bm{\sigma}) \mathbb{I} + \sum_{q=1}^{Q} c'_{q}(\bm{\sigma})| \mathbb{I}_{q} \rangle \! \rangle \otimes | l_{q,1} \rangle \! \rangle,
\end{align}
where $\mathbb{I}_{q}$ refers to the identity operator in all qubit Hilbert spaces except the $q$-th subspace. At this point, we observe that $H^{'}$ takes the local form of Eq. (\ref{Eq:Local_H}) and is Hermitian, because the time evolution of $H$, according to (\ref{Eq:Heis_pict}), can always be written as a sum of time-decaying Hermitian matrices \footnote{It is a direct consequence of the following Liovillian property: if $\rho$ is a left eigenvector of $\mathcal{L}$ (or a right one of $\mathcal{L}^{\dagger}$) with eigenvalue equal to $\lambda$ then  $\rho^{\dagger}$ will be in turn an eigenvector with $\lambda^{*}$ as an eigenvalue. This statement can be proved easily by applying the proof of Lemma 3 in \cite{Spectral_liov} considering the adjoint of a generic Liouvillian.}. Therefore, neglecting terms in the expansion does not preclude the hermiticity of the operator. Consequently, all the results concerning the training of the hyperparameter vector $\bm{\theta}$ for local Hamiltonians can be applied. We emphasize that the condition imposed on the mixing times guarantees that each $\mathcal{L}_q$ operator contributes equally to the expansion of the slowest terms of Eq. \ref{Eq:Approx_Mark}, resulting in the definition of $H'$. It is also easy to verify that the emergence of an effective Hamiltonian with local character holds even when faster modes are included.

Our strategy is effective only if the $c'$ coefficients are not negligible. This property is achieved when the slowest decay right eigenvectors of $\mathcal{L}$ have a significant overlap with $H$ and, consequently, the choice of the Liouvillan has to be designed according to the known Hamiltonian properties. 

 In addition to avoiding the problem of the barren plateau, we have to make sure that the minimum energy of $H^{'}$ and the minimum energy of $H$ are close enough. Such a closeness will depend, in general, on the value of the parameters $\bm{\sigma}$, and, consequently, the trainability of the non-unitary layer is a property to take into account. To this end, we study how the 
derivative of the cost function varies as a function of a generic component of $\bm{\sigma}$, which we call $\sigma_{k}$. 
As in the previous case of the unitary parameters, a key quantity to calculate is the variance of $\sigma_{k}$, which is related to the probability of sampling a non-negligible value of such parameter. If we call $d\mu(\bm{\theta})$ and $d\mu(\bm{\sigma})$ the distribution volume elements of the free parameters, we obtain

\begin{align}
     \Var \left[{\frac{\partial C}{\partial \sigma_k}}\right] &= 
     \Var \left[ \Tr{ \frac{\partial H^{'}}{\partial \sigma_k} \rho(\bm{\theta})} \right  ] \nonumber \\
    &= \Var \left[ \Tr{ \Tilde{H}(\bm{\sigma}) \rho(\bm{\theta})} \right] \nonumber \\
    &= \int_{\bm{\sigma}} d\mu(\bm{\sigma})\int_{\bm{\theta}} d\mu(\bm{\theta}) \left( \Tilde{C} - \left\langle\Tilde{C} \right\rangle \right)^{2} \label{Eq:der_sigma} 
\end{align}

where $\Tilde{H}$ is a local Hamiltonian that can be written in the form of Eq. (\ref{Eq:H'}) and $\Tilde{C}$ is its relative cost function computed with respect to the unitary ansatz considered.

From Eq. (\ref{Eq:der_sigma}) we learn that an exponential suppression of $\Var [{\partial C/\partial \sigma_k}]$), according to \cite{cost_conc}, can occur if and only if the unitary ansatz presents a barren plateau in the optimization of $\Tilde{H}$. Since $\Tilde{H}$ and $H'$ have the same local character, considering a quantum circuit that is not affected by barren plateaus in the case of local Hamiltonians (see Sec. \ref{Sec:VQE_Unitary}) directly excludes the presence of barren plateaus in the non-unitary layer. 

Moreover, we want to remark that the value of $\Delta t$ that ensures the validity of the approximation in Eq. (\ref{Eq:Approx_Mark})  scales efficiently with the number of generators that compose the model, that is, $\mathcal{O}(\log(Q))$, independent on the specific Hamiltonian $H$ of the problem. Furthermore, the condition $Q = \mathcal{O}(\text{poly}(n))$ ensures an efficient logarithmic scaling even in the number of qubits, as required.

Let us also emphasize that the condition of Eq. (\ref{Eq:Approx_Mark}) can be alleviated by including faster decay terms while still preserving a local structure for $H^{'}$. This means that for practical purposes we can choose a value of  $\Delta t$ significantly smaller than the theoretical threshold discussed above.

Finally, our analysis extends seamlessly to considering a convex combination of non-unitary maps generated by a Liouvillian that satisfies the constraints introduced in Sec. \ref{Sec:Non_Un}. Indeed, a non-unitary layer given by the convex combination $\mathcal{E} = \sum_{j} \beta_j e^{\mathcal{L}_j \Delta t}$, in addition to being a suitable quantum channel, will still transform a global Hamiltonian to a local one. Importantly, if we compute the variance of the derivative of the cost function with respect to either the $\beta_j$ coefficients or the $\mathcal{L}_j$ free parameters, we find that the result reduces to the form shown in Eq. \ref{Eq:der_sigma}. This observation generalizes the conclusions drawn from our previous analysis.

\section{Random Hamiltonian example}
\label{Sec:Rand_Ham}
We will now provide a numerical demonstration of the scaling calculated analytically in Sec.~\ref{Sec:main} in a synthetic example. We consider a random Hamiltonian whose ground state in each realization can be either in the neighborhood of the state $\ket{\bm{0}}$ or of  $\ket{\bm{1}}$. Assuming lack of knowledge of the ground state, the initial guess for the ansatz is random. For the sake of definiteness, we have fixed the minimum energy to the value of -1.1 (see Appendix \ref{App:Rand} for more details). Moreover, as in Refs. \cite{mcclean2018barren, expressibility}, we will employ a hardware-efficient, layered quantum circuit as the unitary component of our ansatz (its form is shown in Appendix~\ref{App:QCirc}). Considering the known properties of the Hamiltonian, we define the non-unitary layer as follows:
\begin{align*}
\mathcal{E}(\sigma) = s(\sigma) e^{\mathcal{L}^{(1)} \Delta t} + \left[1-s(\sigma)\right] e^{\mathcal{L}^{(2)} \Delta t},
\end{align*}
where $\sigma$ is a real free parameter, $s(\sigma)$ is the sigmoid function that ensures the sum is a convex combination, and $\mathcal{L}^{(1)}$ and $\mathcal{L}^{(2)}$ are Liouvillians made up of single-qubit dissipators such that their corresponding stationary states are $\ket{\bm{0}} \bra{\bm{0}}$ and $\ket{\bm{1}} \bra{\bm{1}}$, respectively. These two Liouvillians can be easily derived considering the superoperators family identified in Appendix C in such a way that they will respect the needed constraints presented in Sec.~\ref{Sec:Non_Un}. This particular choice ensures a non-negligible overlap with the ground state of the target Hamiltonian. Moreover, by training the parameter $\sigma$, we can transform the outputs of the quantum circuit into states close to the desired ground state. This convex combination can be implemented by considering a stochastic Liouvillian, sampling $\mathcal{L}^{(1)}$ with probability $s(\sigma)$ and $\mathcal{L}^{(2)}$ with complementary probability. 

\begin{figure*}[]
    \includegraphics[width= \linewidth, keepaspectratio]{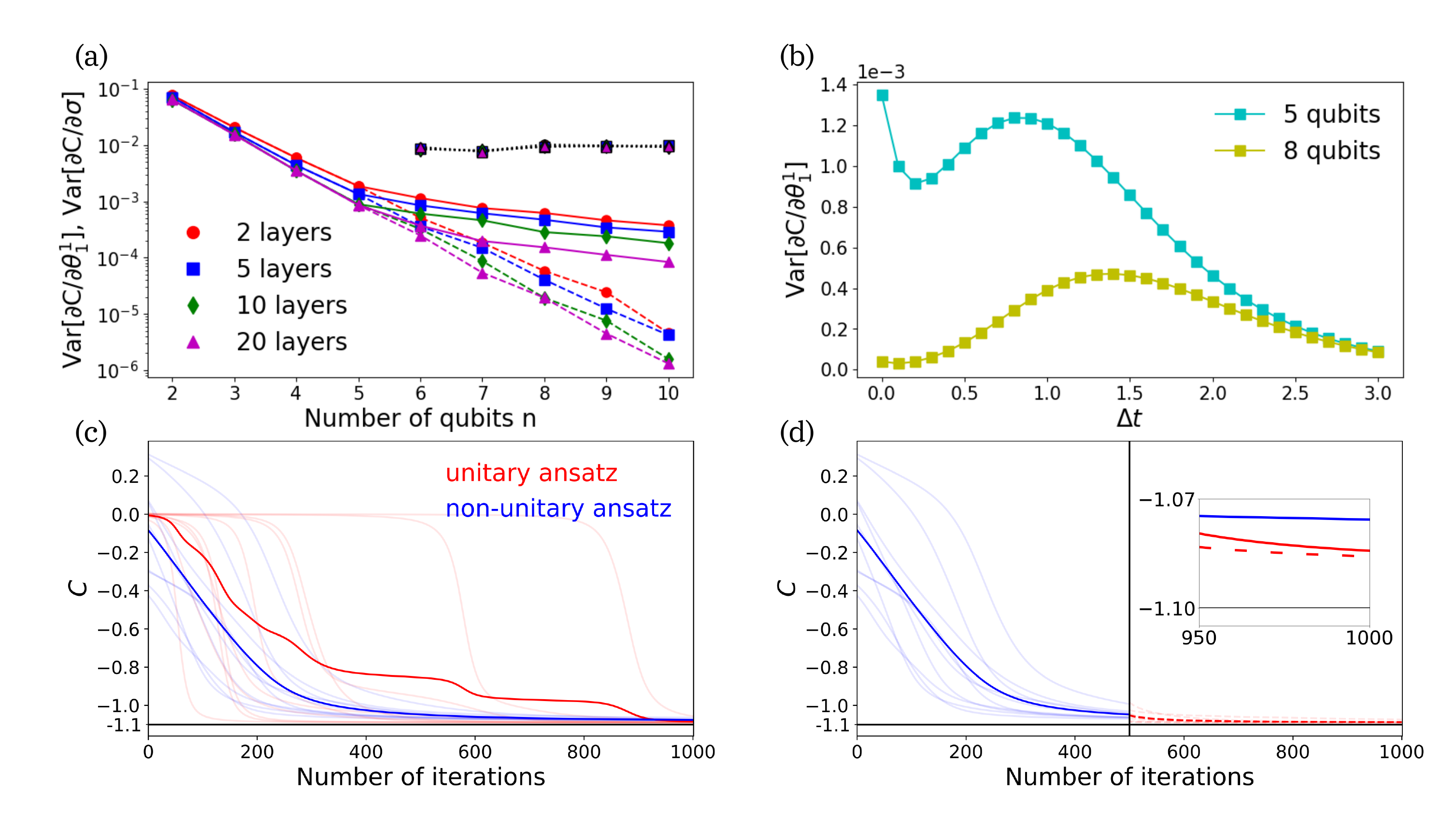}
    \caption{Results of the random Hamiltonian example. (a) Scaling of the partial derivative variances with respect to $\theta_{1}^{1}$ and $\sigma$. Each point was determined from a statistics of $1000$ different random samples. It includes different rotations directions, Hamiltonian and free parameter realizations. Moreover, the $x$ parameter was selected at random in the range $[-5, 5]$ while all the angles were also selected at random in the compact set $[0, 2\pi]$. For the non-unitary ansatz, we varied $\Delta t$ from $0.1$ to $3$ in steps of $0.1$, selecting the biggest variance for each value of $n$. The variances with respect to $\theta_1^1$ are depicted by dashed lines for the fully unitary ansatz whereas, for the non-unitary ansatz case, they are represented by continuous lines. The partial derivative variance with respect to $x$ is depicted using the black dotted lines in the upper part of the plot. We only plot the cases where the non-unitary ansatz is beneficial for training (for very short chains no improvement is observed compared to the unitary ansatz ). (b) Variance averages with respect to $\theta_{1}^{1}$  as a function of $\Delta t$ in the case of a 5-layer quantum circuit, for 5 and 8 qubits.  (c)-(d) Evolution of the cost function evolution under the gradient descent algorithm with a learning rate of $0.1$ and for $10$ random initial conditions. The light lines represent single realizations, while the relative averages are reported in a darker color. (c) The performances of the two different cases (unitary vs non-unitary ansatz) are compared. (d) Cost function calculated through a hybrid approach: the first 500 iterations involve non-unitary ansatzes, while in the remaining 500 we impose the condition $\Delta t = 0$.}
    \label{fig:rand_herm}
\end{figure*}

In our numerical experiment, we first consider computing the derivative variance scaling with respect to the unitary ansatz parameters, with and without the addition of the non-unitary layer. In the case of the non-unitary ansatz, we have spanned $\Delta t$ from $0.1$ to $3$ with a resolution of $0.1$ to find its optimal values. In Fig. \ref{fig:rand_herm}(a) we can clearly observe that, for a fixed number of layers of the quantum circuit, dissipation enables the prevention of an exponential scaling of the gradient variance, which indicates the absence of the barren plateau, as we predicted. 
We also remark that the optimal values of $\Delta t$ found are always below the characteristic time threshold indicated in Sec. \ref{Sec:main}. To give an idea of the role of the dissipation strength, we plot the trend of the variance as a function of $\Delta t$ in Fig. \ref{fig:rand_herm}(b). 
We find a  non-monotonic trend that is the result of the competition between the constructive role of dissipation, leading to an increase of locality, but also the drawbacks of a contractive dynamics. The latter at long time would lead to a full Hamiltonian erasure (large $\Delta t$ in Fig. \ref{fig:rand_herm} 2 (b)), while in the initial transient can rescale the Hamiltonian with a consequent reduction of the variance (more visible for 5-qubits). For the non-unitary layer to provide an advantage, the number of qubits has to be large enough to obtain a maximum that overcomes the initial value, which corresponds to the fully unitary case. In Fig. \ref{fig:rand_herm}(a), we also show the behavior of the derivative of the variance with respect to the free parameter of the unitary layer as the number of qubits is changed. We can clearly exclude the presence of an exponential scaling, in agreement with the theoretical results of Sec. \ref{Sec:main}.

Finally, it is also important to test the accuracy of the ground-state energy estimation for both the unitary and the non-unitary ansatz. To this end, we applied $1000$ iterations of the gradient descent algorithm, with a learning rate of $0.1$, to $10$ random initial conditions in the ideal case where the gradient can be exactly estimated. The results, presented in Fig. \ref{fig:rand_herm}(c), show that the non-unitary circuit allows a faster convergence, with a relative average error of $2.2\%$, while the fully unitary circuit has a slower convergence but a higher accuracy of $1.5\%$. This is expected since the fully unitary circuit preserves the original Hamiltonian and hence its ground state.

To take advantage of both properties, we propose a hybrid approach where the non-unitary ansatz is applied to the initial state for the first $500$ iterations, and then only the fully unitary circuit is used for the last $500$ iterations by setting $\Delta t$ to zero. This method serves as a novel initialization strategy, and from Fig. \ref{fig:rand_herm} (d), we observe a significant improvement in convergence time compared to the fully unitary case. Additionally, the average accuracy is the best found with a value of $1.2\%$.

\section{Quantum chemistry example}\label{Sec:Chem_Ex}

\begin{figure}[!t]
    \includegraphics[width=\linewidth, keepaspectratio]{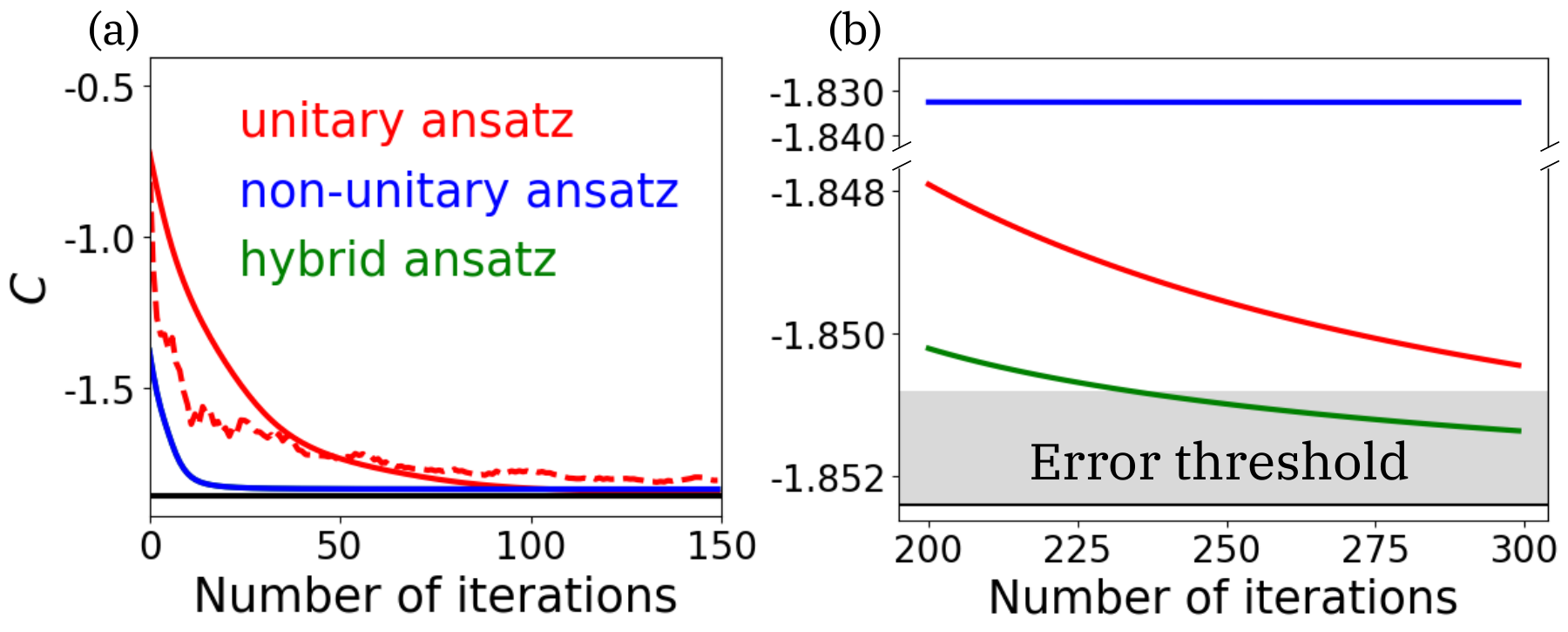}
    \caption{Results of the gradient descent algorithm for the example of $H_2$.  (a) First 150 iterations of the algorithm. The blue line refers to the non-unitary ansatz optimization with $\Delta t = 0.5$ and a learning rate equal to 1, the red dotted one to the unitary-ansatz optimized with a learning rate equal to 1, while, for the red continuous one, the unitary ansatz takes a learning rate equal to 0.1. The vertical black line indicates the ground-state energy computed by a numerical diagonalization of the Hamiltonian. (b) Last 150 iterations. The green line shows the hybrid ansatz optimization and the grey region refers to energy values falling within the considered error threshold. The hybrid ansatz learning rate is set to 0.1.}
    \label{fig:h2}
\end{figure}

Moving to a more practical application, we show the results of the beneficial effects of our non-unitary model in the context of quantum chemistry. In particular, we have studied the problem of finding the electronic ground state of the $H_2$ molecule. In our numerical experiments, we studied the molecule in its equilibrium configuration where the two hydrogen atoms are separated by $0.74$$\textup{~\AA}$. The qubit Hamiltonian $H$ was obtained from a Jordan-Wigner transformation \cite{Jordan1928}, while the electronic orbitals were generated from the STO-3G basis set \cite{sto3g}. Under these conditions, $H$ spans a four-qubit Hilbert space. Following the previous example, the quantum circuit that defines the unitary part of the ansatz is the one presented in Appendix \ref{SubSec:quantum_circ} with a number of layers fixed to $20$. As for the non-unitary layer, the nature of the problem suggests choosing a Liouvillian whose corresponding stationary state is the Hartree-Fock state \cite{HF}. Interestingly, the class of Liouvillians presented in Appendix \ref{App:single_diss} can always accomplish this task by simply dissipating along the $z$ direction, because the Hartree-Fock state is generally a computational basis state. 

As in the previous example of Sec. \ref{Sec:Rand_Ham}, we have applied the gradient descent algorithm to ten random initial conditions both in the case of a fully unitary ansatz, which uniquely consists of a quantum circuit, and in the case of a non-unitary one. We found that a value of $\Delta t = 0.5$ is sufficient to significantly improve the convergence time and the ground state quality as shown in Fig. \ref{fig:h2}. In particular, we have found that in the non-unitary case, the algorithm is able to converge by fixing a learning rate of $1$, while in the fully unitary case, we need to set the learning rate to $0.1$  to achieve faster convergence time and a lower value of the final ground state. In all the considered examples, we have fixed the number of iterations to 300. As in the random Hamiltonian example, we observe that the introduction of the non-unitary layer makes it possible to significantly reduce the algorithm time of almost two orders of magnitude, at the cost of convergence to a less accurate ground state. 

As done in the previous section, we propose a hybrid approach in which, we first apply the non-unitary ansatz, and after half of the total iteration steps, we remove the dissipation and apply only the unitary ansatz to complete the experiment. As shown in Fig. \ref{fig:h2}, this strategy is successful both in reducing the convergence time and in improving the quality of the final ground state. We have set an error threshold with respect to the ground state energy calculated using a numerical diagonalization equal to 0.00159 Hartree, in line with the standard threshold for defining chemical accuracy. Notably, only the hybrid approach can estimate the ground state energy within this energy interval.

\section{Discussion}\label{Sec:conclusions}

Extensive research in the field of variational quantum algorithms is devoted to exploring strategies and techniques aimed at circumventing or minimizing the occurrence of barren plateaus. This is essential for improving scalability, algorithmic efficiency, and applicability in a wide range of contexts. Here we have proposed and demonstrated the effectiveness of a strategy based on engineered dissipation, that can be understood as a localization of a global cost function argument. This dissipation-based analysis goes beyond the general scaling laws recently presented in Refs. \cite{formula_BP1, formula_BP2}, which are limited to the use of a unitary ansatz. Indeed, the presented mitigation strategy has a time complexity that scales logarithmically as the number of qubits is increased, which makes it possible to estimate the ground state energy of systems that would be intractable with previous methods.

To assess the effectiveness of our approach, we have first examined the disappearance of the barren plateau resulting from the presence of engineered dissipation in a simple illustrative model that admits an analytical solution. This toy model also serves as a useful point of comparison for the scenario with general noise, which worsens the problem of attenuation of the barren plateau. Interestingly, handling Hamiltonians such as $H = \mathbb{I} - \ket{\psi}\bra{\psi}$, where $\ket{\psi}$ represents a generic pure state, can be challenging with alternative proposals, such as perturbative gadgets \cite{pert_gadg}. While both these approaches map a global Hamiltonian into a local one,  our engineered dissipation strategy does not require to increase the Hilbert space dimension and the time complexity is entirely unrelated to the decomposition of the Hamiltonian using Pauli matrices.   

The formal theoretical framework of this proposal stands on the analytical proofs in Sec. \ref{Sec:main}, where we first demonstrate that the presence of a tailored non-unitary layer makes it possible to map the cost function into an effective one that corresponds to a local Hamiltonian. Moreover, we also prove that this type of non-unitary layer can be efficiently trained. 
{This theoretical proposal opens a new avenue for implementations in noisy quantum processors. Actually, the form of engineered dissipation here presented can be efficiently implemented on experimental platforms, for instance through collision models \cite{CM_ibm}}, as discussed in Appendix \ref{App:Impl}. 

We then tested the use of nonunitary ansatzes in two relevant numerical examples.  The first one addressed a random synthetic Hamiltonian, while the second tackled the realistic task of determining the lower energy of a Hydrogen molecule. Both problems display the effectiveness of our method in speeding up the convergence and also the precision reached in the estimation of the ground state energy. Indeed, the non-unitary approach can also be seen as an efficient initialization protocol for a unitary ansatz suggesting a powerful hybrid strategy where a unitary ansatz follows a non-unitary one. 

Our results are consistent with other quantum machine learning protocols, where the presence of losses can enhance performance. 
In a recent work, a dissipative optimization algorithm, able to efficiently find Hamiltonian local minima, has been reported \cite{local_m}.
Another relevant example can be found in the realm of quantum reservoir computing \cite{opportunities}, where dissipation can be transformed into a constructive resource \cite{PhysRevResearch.5.023057,sannia2022dissipation,Domingo2023}. To stay within the context of VQAs, it was demonstrated that incorporating stochastic noise can prevent the occurrence of saddle points, which are detrimental to efficient optimization \cite{liu2022noise}. Finally, we proposed a simple dissipation model that proved very effective in mitigating barren plateaus. This indicates the potential for devising more comprehensive strategies that fully explore the potential of non-unitary architectures in variational quantum circuits but also in the broader arena of quantum neural networks.   

\section*{Acknowledgements}
 We acknowledge the Spanish State Research Agency, through the Mar\'ia de Maeztu project CEX2021-001164-M funded by the MCIN/AEI/10.13039/501100011033 and through the QUARESC project (PID2019-109094GB-C21/AEI/10.13039/501100011033), MINECO through the QUANTUM SPAIN project, and EU through the RTRP - NextGenerationEU within the framework of the Digital Spain 2025 Agenda.
We also acknowledge funding by CAIB through the QUAREC project (PRD2018/47). The CSIC Interdisciplinary Thematic Platform (PTI) on Quantum Technologies in Spain is also acknowledged. GLG is funded by the Spanish  Ministerio de Educaci\'on y Formaci\'on Profesional/Ministerio de Universidades and co-funded by the University of the Balearic Islands through the Beatriz Galindo program (BG20/00085). The project that gave rise to these results received the support of a fellowship from the ”la Caixa” Foundation (ID 100010434). The fellowship code is LCF/BQ/DI23/11990081.
IBM, the IBM logo, and ibm.com are trademarks of International Business Machines Corp., registered in many jurisdictions worldwide. Other product and service names might be trademarks of IBM or other companies. The current list of IBM trademarks is available at \url{https://www.ibm.com/legal/copytrade}.

 \appendix
\section{Random Hamiltonian structure}\label{App:Rand}
In Sec. \ref{Sec:Rand_Ham}, we have proposed the optimization of a random Hamiltonian $H$ from which we have only partial information about a possible ground state. We will now show the criteria used to randomly generate $H$ in such a way that it is constrained to this incomplete ground state knowledge. We point out that the non-unitary optimization algorithm shown in Sec. \ref{Sec:Rand_Ham} only has access to this Hamiltonian property and, as a consequence, the key idea can be applied in a general context. 

Going now into the details of $H$, for the sake of definiteness, we will impose the condition that all its eigenvalues $\lambda_i$ satisfy $|\lambda_i| < 1$, except for the maximum eigenvalue which is fixed at 1.1, and the minimum eigenvalue which is set at -1.1. Additionally, we have constrained the eigenvectors associated with these last two values to be one of two states (up to a normalization factor): $\ket{\psi_1} = \ket{\bm{0}} + 0.1 \ket{\phi_1}$ and $\ket{\psi_2} = \ket{\bm{1}} + 0.1 \ket{\phi_2}$, selected randomly and exclusively, where $\ket{\phi_1}$ and $\ket{\phi_2}$ are Haar-distributed states. We stress that for $\ket{\psi_1}$ and $\ket{\psi_2}$ to be, in general, viable eigenvectors, they must undergo a Gram-Schmidt orthonormalization.

Finally, we underline that the algorithm of Sec. \ref{Sec:Rand_Ham} is built from two possible ground-state neighborhood guessing which are referred to two opposite cases: the maximum and the minimum eigenvalue. Consequently, the numerical results obtained indicate that the algorithm was able to successfully discriminate them.
\section{Quantum circuit}\label{App:QCirc}
\begin{figure*}[!t]
    \includegraphics[width = 0.6 \linewidth, keepaspectratio]{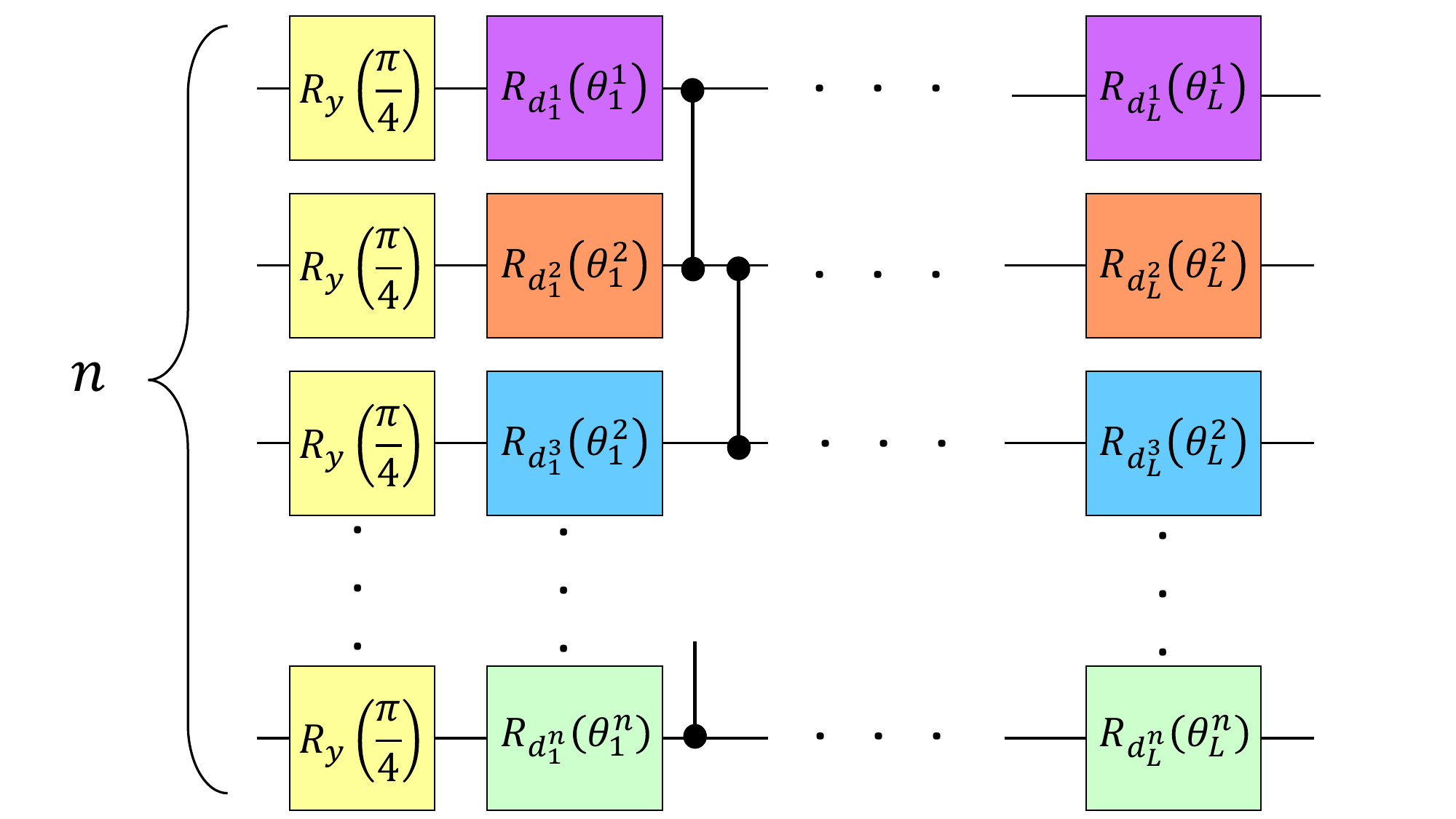}
    \caption{Quantum circuit employed in the numerical simulations of Sections \ref{Sec:Rand_Ham} and \ref{Sec:Chem_Ex}.}
    \label{fig:circ}
\end{figure*}
\label{SubSec:quantum_circ}
We now show in detail the form of the quantum circuit considered in Sections \ref{Sec:Rand_Ham} and \ref{Sec:Chem_Ex}. Following the refs. \cite{mcclean2018barren, expressibility}, it has the structure of a hardware-efficient, layered quantum circuit  structured as follows:
\begin{equation*}
U(\bm{\theta},\bm{d}) = \prod_{l=1}^{D} W U_l(\bm{\theta}_l,\bm{d}_l).
\end{equation*}
This comprises $D$ layers of rotations and entangling gates, with the entangler $W$ consisting of a series of controlled-$Z$ gates that correlate adjacent qubits:
\begin{equation*}
W = \prod_{i=1}^{n-1} CZ_{i,i+1}.
\end{equation*}
A $U_l$ gate, on the other hand, is formed by single-qubit rotations in random directions:
\begin{equation*}
U_l(\bm{\theta}_l,\bm{d}_l) = \prod_{i=1}^{n} R_{d_l^i}(\theta_l^i),
\end{equation*}
where $R_{d_l^i}(\theta_l^i)$ is a rotation gate that applies an angle of $\theta_l^i$ around the  direction $d_l^i$ to the $i$-th qubit, and $d_l^i$ is randomly chosen from the $x$, $y$, or $z$ axis, while $\theta_l^i$ takes values in the set $[0, 2\pi]$. To avoid any preferential direction, the qubits are initially prepared in the state $\rho_{in} = \ket{\psi_0}\bra{\psi_0}^{\otimes n}$, where $\ket{\psi_0} = R_y(\frac{\pi}{4}) \ket{0}$. The all circuit structure is summarized in fig. \ref{fig:circ}.

\section{Liouvillians based on single-qubit dissipators }\label{App:single_diss}

We now present a simple class of Liouvillians that satisfy the constraints required for our purposes (see Sec. \ref{Sec:Non_Un}) from which we have built the non-unitary layer considered in sections \ref{Sec:Rand_Ham} and \ref{Sec:Chem_Ex}. We consider a generator $\mathcal{L}$ that takes the form of Eq. (\ref{Eq:local_dec}), where each operator $\mathcal{L}_q$ implements single-qubit dissipation along a desired direction. More precisely, their action is described by the following relation:

\begin{align}
\mathcal{L}_q \rho \nonumber = d_{q} \rho d_{q}^{\dagger}-\frac{1}{2} \{d_{q}^{\dagger}d_{q}, \rho \} ,
\end{align}\label{eq:Model_Prop}
where $d_q$ is a jump operator acting onto the q-th qubit Hilbert space:

\begin{align*}
d_q = \ket{\psi_- (\alpha_{q}, \phi_{q})}_q \bra{\psi_+ (\alpha_{q}, \phi_{q})},
\end{align*}
with

\begin{align*}
 \ket{\psi_+ (\alpha_{q}, \phi_{q})} &= \cos(\frac{\alpha_{q}}{2}) \ket{0} + e^{i \phi_{q}} \sin(\frac{\alpha_{q}}{2}) \ket{1}, \\
 \ket{\psi_- (\alpha_{q}, \phi_{q})} &= \sin(\frac{\alpha_{q}}{2}) \ket{0} - e^{i \phi_{q}} \cos(\frac{\alpha_{q}}{2}) \ket{1}. 
\end{align*}
Here the parameters $\alpha_q$ and $\phi_q$ identify the direction of the dissipation. This choice satisfies the required in Sec. \ref{Sec:Non_Un} because the spectral gap of each $\mathcal{L}_q$ is identically equal to $1/2$. Moreover, each $\mathcal{L}_q$ has as a unique stationary state $\ket{\psi_- (\alpha_{q}, \phi_{q})}_q \bra{\psi_- (\alpha_{q}, \phi_{q})}$, and this information can be used to study the overlap with the target Hamiltonian to ensure the effectiveness of the approach.

\section{Non-unitary layer implementation}\label{App:Impl}

Nonunitary operations can be effectively realised on digital quantum computing architectures in several ways. This task is in fact closely related to the general problem of implementing quantum simulation algorithms for open systems~\cite{Miessen2023,kliesch_dissipative_2011,endo_variational_2020,kamakari2021digital,Ramusat2021quantumalgorithm}. The most direct approaches employ, for instance, dilation theorems allowing one to recast dissipative dynamics into a unitary evolution on an extended system~\cite{cleve2017efficient,khm2021capturing,CM_efficient}. Alternatively, or in combination with the latter, one could in principle make use of engineered dissipations leveraging intrinsic qubit noise~\cite{verstraete2009quantum,Murch2012,rost2021demonstrating,leppakangas2022quantum}, randomized schemes~\cite{chenu2017,pec2023} or controlled classical environments~\cite{potocnik_studying_2018}.

To give a concrete example, we now present an efficient implementation of the nonunitary layer by means of the so-called collision models (CMs) \cite{CM_review}. The general idea of CMs is to approximate the Markovian dynamics given by Eq. (\ref{eq:Linb}) by letting the system qubits sequentially interact in a unitary way with a set of ancillary qubits. In particular, the following approximation is considered:
\begin{align*}
     e^{\mathcal{L}\Delta t} \approx (\phi_{\Delta t /M })^M
\end{align*}
which consists of the alternating M steps where a quantum channel $\phi_{\Delta t /M }$ approximates the evolution for a finite time equal to $\Delta t / M$. By definition,
\begin{align}\label{Eq:CM_step}
    \phi_{\Delta t /M } [\rho] = \Tr_a  \{ V(\Delta t /M) \rho \otimes \rho_a V^{\dagger}(\Delta t /M )\} 
\end{align}
where $\Tr_a$ indicates the partial trace over the ancillary qubits space, $\rho_a$ is their state which has to be properly prepared at each step and $V(\Delta t /M)$ is a unitary operator which, depending on the particular dynamics of interest, non-trivially acts on the system and ancillary qubits space. 

Importantly, for local Liouvillians as required by the criteria of Sec. \ref{Sec:Non_Un}, the number of resources needed by the collision-model strategy always scales efficiently \cite{CM_efficient}. 
More quantitatively, 
let us call $\epsilon$ the difference between the exact Liouvillan dynamics and the $M$-step collision model:
\begin{equation*}
    \| e^{\mathcal{L}\Delta t} - (\phi_{\Delta t /M })^M  \|_{1 \rightarrow 1} = \epsilon,
 \end{equation*}
 where $\left \| \cdot \right \|_{1 \rightarrow 1}$ indicates the trace norm. Then, it can be shown that the number of gates necessary to have an error $\epsilon$  is always a polynomial function of the number of qubits $n$.

According to Ref.~\cite{CM_ibm}, the class of models defined in Appendix Sec.~\ref{App:single_diss} can be efficiently implemented by coupling each qubit with an ancillary one initialized in the state $\ket{0}$. Going more into details, in the definition of $\phi_{\Delta t /M}$, according to Eq. \ref{Eq:CM_step}, we will have $\rho_a = \ket{0}\bra{0}^{\otimes n}$ and  $V(\Delta t /M) = \prod_{q = 1}^{n} V_q(\Delta t /M)$ with 
\begin{align*}
    V_q(\Delta t /M) = \exp(-i \sqrt{\Delta t /M}(d_q \sigma^{+}_{q^{(a)}} + h.c. ) )
\end{align*}
where the index $q^{(a)}$ refers to the q-th ancillary qubit. In view of a possible experimental implementation, we emphasize as a positive note that each unitary $V_q(\Delta t /M)$ can be easily realized through a single-qubit gate and a CNOT gate and that the qubit reset operation, currently available on IBM Quantum devices \cite{Qiskit}, allows the number of ancillary qubits to be fixed to $n$  for the entire execution of the collision-model algorithm.

\bibliography{bibliography}
\end{document}